\RequirePackage[2020-02-02]{latexrelease}

\documentclass[article,twocolumn,amsmath,amssymb,10pt]{revtex4}
\usepackage[pdftex]{graphicx} 
\usepackage{color}

\begin{document}



\title{Transient population dynamics of nanoparticles during pulsed EUV exposures}

\author{M. Chaudhuri$^{1*}$, L. C. J. Heijmans$^1$, A. M. Yakunin$^1$,  P. Krainov$^2$, D. Astakhov$^2$ and M. van de Kerkhof$^{1,3}$}
\affiliation{$^1$ASML The Netherlands B.V., P.O.Box 324, 5500AH Veldhoven, The Netherlands}
\affiliation{$^2$ISTEQ B.V., Eindhoven, The Netherlands}
\affiliation{$^3$Department of Applied Physics, Eindhoven University of Technology, PO Box 513, 5600 MB Eindhoven, The Netherlands}

\begin{abstract}

The transient population dynamics of charged (postive or negative) and neutral nanoparticles have been investigated in a pulsed Extreme Ultra-Violet (EUV) exposure environment with 3DPIC simulations. At the initial stage of the simulation, all the particles are kept neutral. As the number of EUV pulses increases over time, the population of neutral particle decreases faster at the expense of negatively charged particle generation outside the beam location. However, a small population ($< 1\%$) of neutral particles become positively charged due to EUV photon interaction within the beam area and remains in steady state over time. The critical pulse numbers have been estimated for different nanometer size particles above which most of the particles outside the beam locations become negatively charged: smaller is the particle size, larger is the critical pulse number.      
\end{abstract}

\maketitle

Population dynamics in generic terminology is used to model or study the size and lifetime composition of populations as dynamical systems. The size of the population under consideration and the sizes of other interacting populations are correlated with the environmental parameters. The evolution of natural populations is influenced by its interactions with varying surrounding environmental conditions and have an important impact on population dynamics. The models in population dynamics theory have been based on a fundamental principle which says that the relative growth rate of a population is a function of the environment in which it lives. It can be described as the change of population which is the result of four key parameters, such as birth, death, immigration and emigration. These factors can be used to follow and predict changes in a population. Several models have been developed to address population response to a periodically or stochastically changing environment in time~\cite{PD.Taitelbaum.2020}. It is an active research field in climate change~\cite{PD.Taitelbaum.2020}, mathematical biology~\cite{Murray_math_bio}, epidemiology~\cite{PD.Juliano.2007}, ecology~\cite{PD.Chisholm.2014, PD.Kalyuzhny.2015}, genetics~\cite{PD.Beaumont.2009, PD.Assaf.2013}, cosmology~\cite{PD.Weygaert.2011, PD.Avelino.2016}, space debris~\cite{PD.Jurkiewicz.2024}, nonlinear population dynamics~\cite{PD.Adachi.2022}, quantum manifestation of population dynamics~\cite{PD.Blanchet.2015}, economics~\cite{PD.Peterson.2017}, etc. The physics of particle contamination within EUV lithography scanners in the context of semiconductor technology is the latest addition to this fascinating interdisciplinary field due to interactions of nano-/micro-particles with surrounding pulsed EUV and EUV induced plasma environments.

EUV lithography technology is introduced to continue with Moore's law~\cite{Moore.1975} in semiconductor chip manufacturing industry which uses highly energetic EUV photons (energy $\sim$ 92 eV) with the wavelength of 13.5 nm~\cite{Bakshi.2018}. One of the side effect of this development is the generation of EUV induced plasma due to the interaction of such highly energetic EUV photons with low pressure (1-10 Pa) background hydrogen gas~\cite{BanineJAP2006, van_der_Horst_2014, Roderik_2018, Kerkhof_2019, Kerkhof_2020, Kerkhof_2021a,Kerkhof_2021,Kerkhof_2022}. The spatial and temporal evolution of the EUV plasmas has been investigated in laboratory experiments and with particle-in-cell (PIC) models~\cite{van_der_Horst_2016, Astakhov_2016,Beckers_2016,van_de_ven_2018}. The technological complexity increases with pulsed mode operation in complex geometry during which EUV pulse is fired for few tens of nanoseconds duration and creates highly transient non-LTE (Local Thermodynamic Equilibrium) plasma which decays for approx $\sim$ 20 $\mu$s before the next EUV pulse triggers. The major impacts of this developmemt are associated with plasma-surface interaction (critical surfaces in the optical path or nearby area) and plasma-particle interactions.
Whenever the plasma comes in contact with surfaces, it creates a local electric field which depends on surface properties (material, sharp edge, roughness, etc). On the other side, the particles get charged when they come in contact with plasma and start responding to the electric field. So, the plasma plays a significant role on the dynamics of mirco- and nanoparticles in the complex geometries of EUV lithographic tools. Understanding and controlling the dynamics of particles are essential for high volume manufacturing of chip production in EUV lithography processes. Even the presence of a single particle on the mask is a high risk as it creates a defect in the imaged pattern on the wafer and all the chips become damaged~\cite{Kerkhof_2019_spie}. 
 Different particle charging mechanisms have been explored in the context of plasma-particle interactions depending on their locations, whether they are at surface~\cite{Cui_1994,Flanagan_2006,Sheridan_2011,Wang_2016,Heijman_2016,
Krainov_2020,Jose_2024,Jose_2025} or in the bulk~\cite{Allen_1957,Allen_1992,Lampe_2001,Lampe_PRL_2001,Kennedy_2002,
Bryant_2003,Hutchinson_2003,Lampe_2003,Ratynskaia_2004,Ratynskaia_2005,
Khrapak_2006,Hutchinson_2007,Khrapak_2008,chaudhuri2010_IEEE,Khrapak2012EPL,Khrapak_PRE_2013}. In recent times, the micro-/nano-particle response to afterglow (both spatial and temporal) physics in the bulk has been studied extensively ~\cite{Ivlev2003_afterglow,Couedel2006,Couedel2008,Couedel2008b,Couedel2009a,Layden2011a,
Denysenko2011,Denysenko2013,Woerner2013,Meyer2016,Merlino2016,
Minderhout2019, chaudhuri_2021,Minderhout2020, Minderhout2021, Minderhout2021_Nat, Denysenko2021, Chaubey2021, Suresh2021, chaudhuri_2022, Chaubey2022, Staps2022, Couedel2022, Huijstee2022}. In the context of pulsed mode operation of EUV lithography scanners, the bulk particle charging mechanisms in highly transient EUV and EUV induced plasma have been explored~\cite{chaudhuri.PSST.2023}. It was found that local transient plasma conditions strongly influence the charge state of the particles: positive, negative or neutral and hence their interactions with local electric field leading to transport from one location to another. In this way, the population of the particles with different charge states are interconnected with surrounding local plasma environments. It is extremely important to control the population dynamics of particles efficiently to minimize contamination risks. {\it The goal of this work is to make the first attempt to explore transient population dynamics of nano particles in pulsed EUV and EUV induced plasmas in such EUV lithographic machines.}

To start with a simplistic approach within our model, the particles are considered conductors, monodisperse and spherical in shape. There are very small number of dust particles in the plasma environment so that local quasi-neutrality condition between electrons and ions are valid.
Depending on the bulk locations (inside or outside of EUV beam) and time (during or after the EUV pulse), the particles are charged through a balance between electron, ion and photon fluxes. The charging equation can be written as,
\begin{equation}
\label{electron_flux} 
\frac{dZ}{dt} = \sum I_\alpha
\end{equation}
Here $Z$ is the particle charge and $I$ is the flux for different plasma components $\alpha$ (= electrons [e], ions [i] or photons [ph]). In our model, the electron flux expression can be written as:

\begin{equation}
\label{electron_flux} 
\begin{aligned}
I_e = \pi a^2 n_e  \times 
    \int\sqrt\frac{2E_e}{m_e}(1-Y(\phi_p, E_e))f_e(E_e) \\ 
\times \left(1 + \frac{e\phi_p}{E_e}\right)\Theta(E_e + e\phi_p) d E_e.
\end{aligned}
\end{equation}
Here, $I_{e},  n_{e}, E_e, m_e$ are the flux, density, energy and mass for electrons respectively. The parameter $f_e(E)$ is the electron energy distribution function, $Y$ is the total yield of secondary electrons and $\Theta$ is a step function. The floating potential of the particle is represented as $\phi_p$ and $a$ is radius of the particle. In detail, the parameter $f_e(E_e)$ represents the flux of electrons with energy $E_e$ to particle surface, $Y(\phi_p, E_e)*f_e(E_e)$ is the flux of secondary electrons from particle surface released by incoming electrons with energy $E_e$. The factor $(1 + e\phi_p/E_e)$ is the collection cross section. Charged particle distorts trajectories of incoming electrons. Positively charged particle attracts electrons with larger impact parameter and vice versa. Electrons with energy $E_e$ less than $|e*\phi_p|$ cannot reach particle surface with potential $\phi_p$ (if $\phi_p < 0)$. Step function theta removes this electrons from integration procedure. In contrast, the ion flux can be written asssuming a thermal distribution of ions with a temperature $T_i$, 
\begin{equation}
\label{ion_flux} 
I_i = \pi a^2\sum_i n_i(q_i/e) v_{Ti} \times \begin{cases}
    \exp(-e\phi_p/kT_i), & \text{if $\phi_p \ge 0$}.\\
    (1 - e\phi_p/kT_i), & \text{if $\phi_p < 0$}.
  \end{cases}
\end{equation}
Here, $I_i,  n_i$ are the flux and density for particular ions respectively. $k$ is the Boltzmann constant and $e$ is the elementary charge. The ion thermal velocity is defined as, $v_{Ti} = \sqrt{kT_i/m_i}$ where $m_i$ is the ion mass.

Similarly, the photon flux to the particle can be written as, 
\begin{equation}
\label{photoelectron_flux} 
I_{ph} = \pi a^2 \gamma(\phi_p) \frac{I}{h\nu}
\end{equation}
Here, $\gamma$ is the photoionization yield (mean number of electrons expelled due to single photon absorption), $I$ is the local EUV intensity at the particle. The correction of a yield can be done by following factor,  
\begin{equation}
\label{photoelectron_flux_correction} 
\gamma(\phi_p) = \gamma_0\int_{e\phi_p}^{+\infty}f_{ph}(E)dE
\end{equation}
Here, $\gamma_0$ is the yield in absence of charge on the particle and $f_{ph}$ is the energy distribution function of the emitted electrons normalized by one.

\begin{figure}[h]
\includegraphics[width=0.7\linewidth]{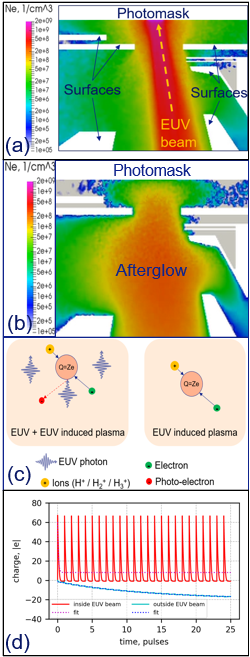}
\caption{(a) Spatial distribution profile of electron density at Reticle Mini-Environment (RME)~\cite{Kerkhof_2021a, Kerkhof_2022} when EUV power is ON for $<$ 100 ns within a single EUV pulse of 20 $\mu$s duration. The yellow dash arrow indicates the EUV-beam path. (b) The afterglow phase at the end of 20 $\mu$s EUV pulse when EUV power is OFF. (c) Two distinct types of particle charging mechanisms which depend on particle location within RME and EUV pulse timing (see text) (d) The charge variation with time (/pulse) for 100 nm particle is shown at the location inside and outside EUV beam.} 
\label{charging_mechanisms}  
\end{figure}

\begin{figure}[h]
\includegraphics[width=0.9\linewidth]{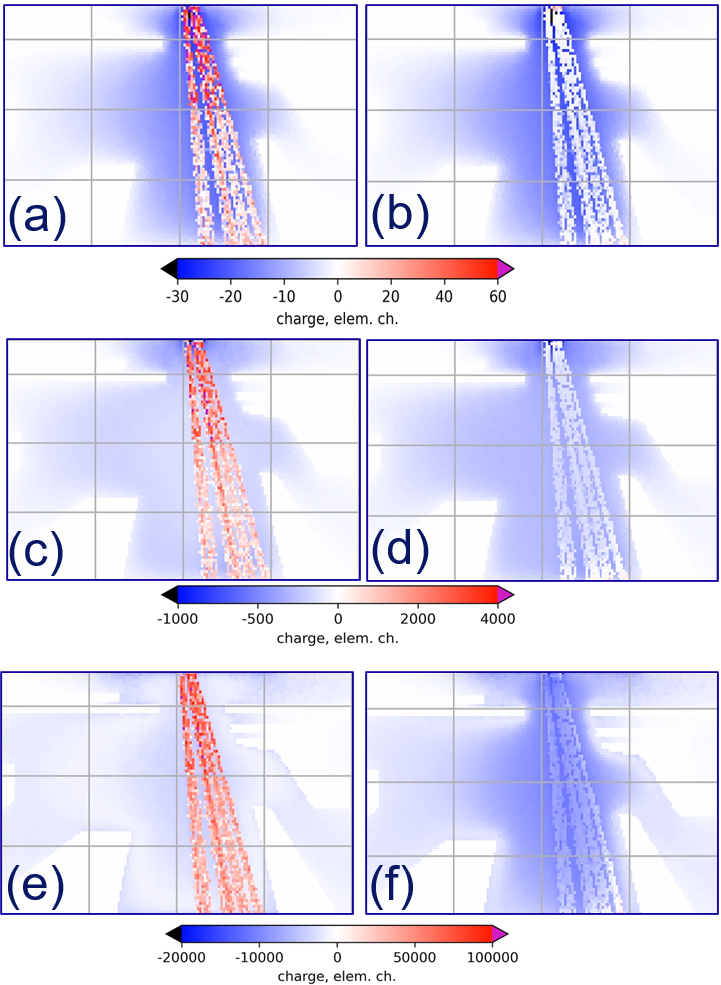}
\caption{Typical population characteristics of charged (postive: red and negative: blue) and neutral (white) particles are shown in the two regimes mentioned in Figure~\ref{charging_mechanisms}a (EUV power ON) and Figure~\ref{charging_mechanisms}b (at the end of 20 $\mu$s EUV pulse when EUV power is OFF) for different particle sizes: 100 nm (a-b), 1 $\mu$m (c-d) and 10 $\mu$m (e-f).} 
\label{transient_plasma_variation}  
\end{figure}

Initially the charge state of all the particles located in the bulk region are considered neutral. As the EUV pulse propageates through background hydrogen gas, the EUV induced plasma is formed. The spatial-temporal evolution of electron density in EUV induced plasma at the beginning and end of EUV pulse is shown in Figure~\ref{charging_mechanisms}(a-b).Then the particles start to interact with surrounding plasma and gets charged or remains neutral which has spatial and temporal dependency. Two distinct types of particle charging mechanisms which depend on particle location within RME and EUV pulse timing as shown in Figure~\ref{charging_mechanisms}c: (1) the EUV photon flux to the particle plays important role along with electrons and ions flux when the particles are within EUV beam path and EUV pwer ON. (2) the traditional charging mechanism (balance between electron and ion fluxes from EUV induced plasmas) occurs for particles located outside EUV beam path when EUV power is ON and when EUV power is OFF. The charge variation of 100 nm particle inside and outside EUV beam in the bulk is shown in Figure~\ref{charging_mechanisms}d. Furthermore, different size particles interact differently with the same background plasma. The spatial-temporal evolution of charging-decharging dynamics of 100 nm, 1 $\mu$m and 10 $\mu$m particles are shown in Figure~\ref{transient_plasma_variation}. Within the EUV pulse, the positive charges acquired by the 100 nm particles are significantly less than that of 1 $\mu$m and 10 $\mu$m particles. Unline 10 $\mu$m particles, the positive charge for 100 nm particles decay faster so that many particles becomes neutral in this phase [Figure~\ref{transient_plasma_variation}(a,c,e) for comparison]. In this phase it is also clear that bigger size particles acquire negative charge much faster than smaller size particles which is reflected in the population of negatively charged particles. The above chacteristics of particles charge dependence on size gets even more clarity in the afterglow phase as shown in Figure~\ref{transient_plasma_variation}(b,d,f).

\begin{figure}
\includegraphics[width=0.9\linewidth]{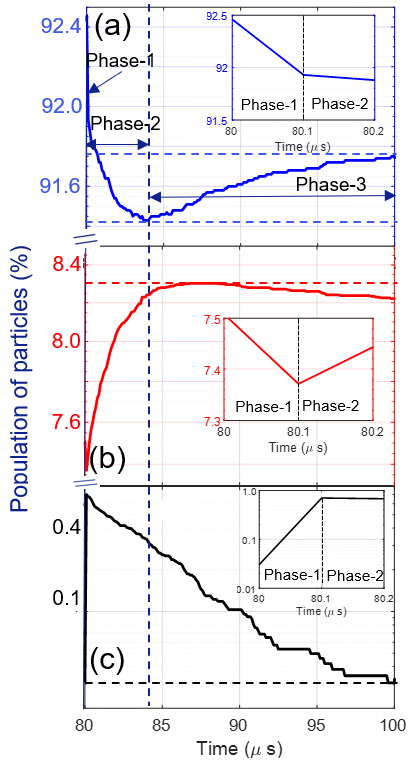}
\caption{The detailed transient population dynamics of (a) neutral particles (blue), (b) negatively charged particles (red) and (c) positively charged particles (black) within single pulse (Pulse$\#$04: 80-100 $\mu$s) with pulse energy 0.2 mJ. The particle size is 100 nm. The qualitative features of such single pulse transient dynamics of particle populations are generic for all pulses. The transient population dynamics consists of three phases as shown in the figure. Phase-1 corresponds to the EUV-ON case and remaining phases correspond to the EUV-OFF cases. The qualitative changes of population dynamics between phase-1 and phase-2 is shown in the inset. The detailed description is mentioned in the text. }
\label{transient_population_dynamics_1pulse}  
\end{figure}

\begin{figure}
\includegraphics[width=0.9\linewidth]{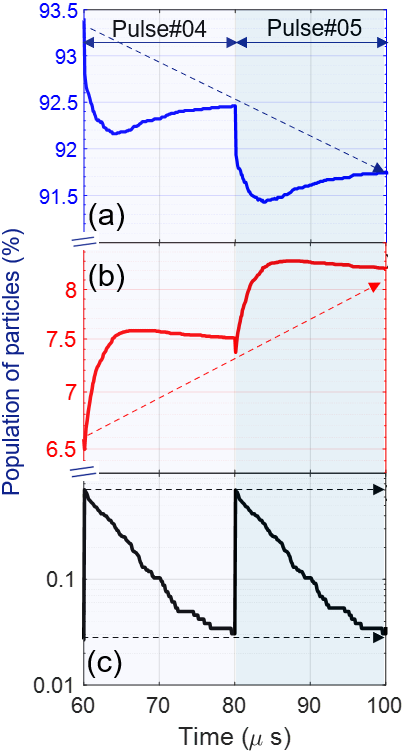}
\caption{Transient population dynamics of (a) neutral particles (blue), (b) negatively charged particles (red) and (c) positively charged particles (black) for 4th and 5th pulses. The particle size is 100 nm. The average population of neutral particles decreases fast at the expense of negatively charged particle generation. However, the average population of positively charged particles remains constant over time. The arrows represents illustration of the population dynamics features. The detailed description is mentioned in the text.}
\label{transient_population_dynamics_2pulses}  
\end{figure}

The transient population dynamics of neutral particles as well as different polarity (positive and negative) charged particles within a single pulse have been investigated in detail which is shown in Figure~\ref{transient_population_dynamics_1pulse}. Considering 20 $\mu$s single pulse duration, the complete transient population dynamics within each pulse can be divided in three phases with different time scales: The phase-1 exists during EUV power ON mode ($<$ 100 ns) during which the population of neutral particles declines sharply ($\sim 0.5\%$). The interesting dynamics is associated with population of negatively charged particles which declines up to 100 ns ($\sim 0.16\%$) and then starts increasing. This is due to the fact that the negatively charged particles from the previous pulse are hit by the EUV photons within 100 ns which converts them to positively charged particles. As a result of such conversion, the steepest increase of population of positively charged particles is observed ($\sim 0.9\%$). Phase - 1 is the shortest period which is followed by phase - 2 which exists between 100 ns and 4$\mu$s. At the beginning of phase - 2 ($< 100 {\rm ns}$), the population of neutral particles decays at very slow rate (compared to phase - 1) whereas that of negatively charged particles sharply increases and return back to its initial value at the beginning of pulse. In this time period the population of positively charged particles almost remains same. After this initial time scale within phase - 2, the population of positively charged particles decreases steadily due to fast charge reduction and they immigrate to the population of neutral partices. But outside the EUV beam path, the population of negatively charged particles increases shaprly at the expense of neutral particles population. At the end of phase-2, the population of nagatively charged particle reaches maximum and that of neutral particles reaches minimum. After completion of phase - 2, the phase - 3 period triggers which is the longest period and lies between 4$\mu$s and 20$\mu$s. In this phase almost all the positively charged particles lose their charge and dynamic immigration happens from population of positively charged particles towards neutral particles. This is evident from the population plots of neutral and positively charged particles. On the other hand, in this phase the population of negatively charged particles almost remains in saturation for $\sim$ 6 $\mu$s and then slowly starts to decline due to conversion from negative to neutral particles as observed in Figure~\ref{transient_population_dynamics_1pulse}. During such fast transient processes, the fluctutaions of population dynamics can also be impacted by single particle charge fluctuations. This impact is taken into account with simplistic assumption with a fixed charge fluctuation window between -5e to +5e for neutral particles. Particles with charge above +5e are considered positive and those with below -5e are considered negative.    

\begin{figure}
\includegraphics[width=\linewidth]{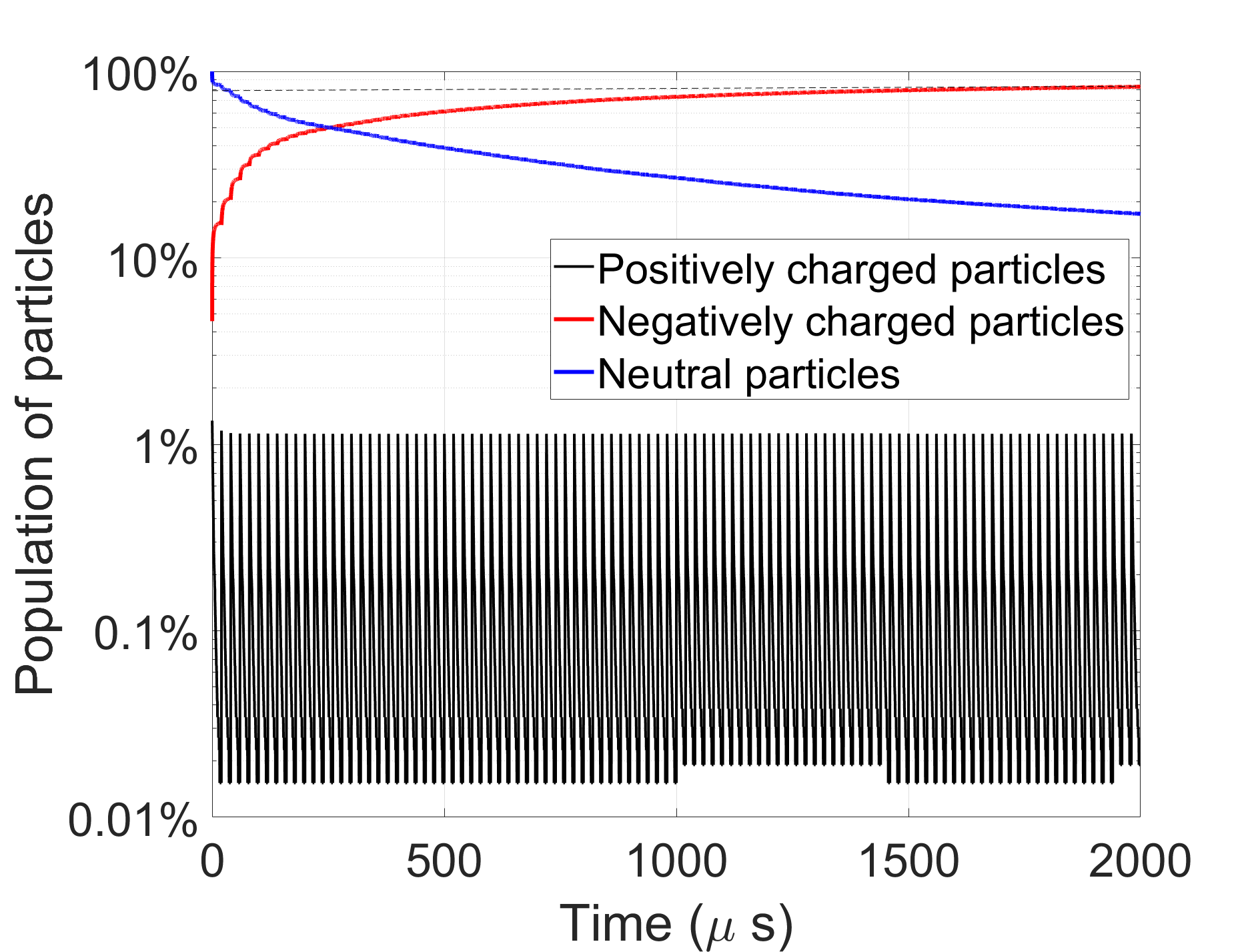}
\caption{Transient population dynamics of (a) neutral particles (blue), (b) negatively charged particles (red) and (c) positively charged particles (black) for 100 pulses. The population variation of positively charged particles remains constant over time. However, the population of neutral particles decreases steadily and that of negatively charged particle increaes. The cross-over pulse number (n$_{\rm co}$) is visible at the intersection point of the population curves for negatively charged particles and neutral particles. The dash line shows the large time scale asymptotic fit for the population of negatively charged particles. The size of the particle is 500 nm. }
\label{transient_population_dynamics_100_pulses}  
\end{figure}

\begin{figure}
\includegraphics[width=0.95\linewidth]{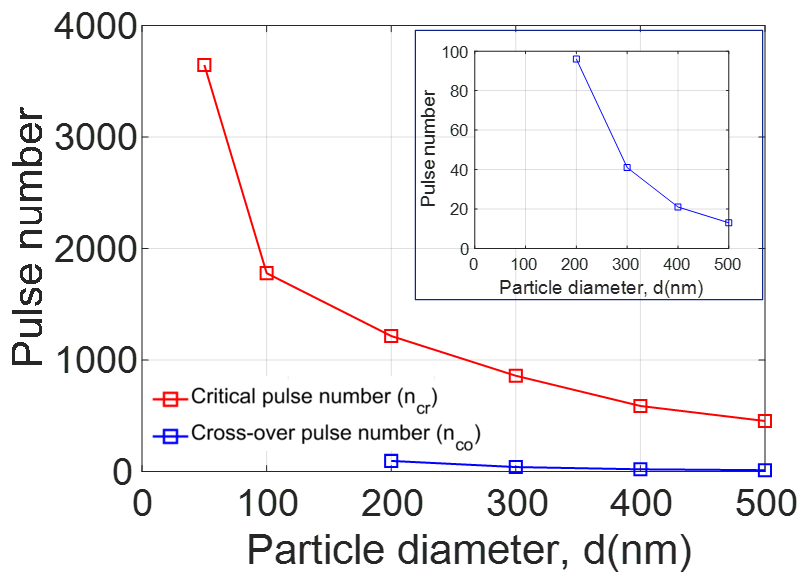}
\caption{The critical pulse number (n$_{\rm cr}$) is estimated for a particular size of particles when all the neutral particles will be converted to negatively charged particles. The variation of such critical pulse number with particle size is shown in this figure (red curve). The smaller the size of the particles, the larger is the critical pulse number. The variation of cross-over pulse number (n$_{\rm co}$) with particle size is shown by the blue curve (also in inset). No cross-over is found for the particles $>$ 200 nm diameter within our simulation range of 100 pulses.}
\label{critical_pulse_number_variation}  
\end{figure}

To get better insight of population dynamics development, the single-pulse scenario is extended for 2-pulses scenario (4th and 5th pulses) as shown in Figure~\ref{transient_population_dynamics_2pulses}. During this time period, the population of neutral particle decreases from 93.4$\%$ to 91.7$\%$ at the expense of steady population rise of negatively charged particles of same magnitude from 6.5$\%$ to 8.2$\%$. Although the population of positively charged particles varies within a single pulse, but such variation remains steady over multi-pulse time frames. Also it is to be noted that the maximum population of positively charged particle reaches only $\sim 0.5\%$ of total population. To get detailed insight of population dynamics for large time scale, 100 pulses (= 2 ms) simulation has been performed as shown in Figure~\ref{transient_population_dynamics_100_pulses}. Although the population of positively charged particles fluctuate within single pulse, but their population shows staeady state over multi-pulse scenario as is also evident from 2-pulses analysis. However, it is clear that the population of negatively charged particles increases steadily over time at large time scale and that of neutral particles decreases. A cross-over point at a certain pulse number can be obtained between the population curve of negatively charged particles and that of neutral particles. Such cross-over point is size dependent which implies that the growth/decay rates of such population curves are size dependent. The larger the size of the particles, the smaller is the cross-over time. As an example, the cross-over pulse number (n$_{\rm co}$) is 12 for 500 nm particle size where as it is extremely large for smaller size particles of 10nm. To make an estimation for the critical pulse number ($n_{\rm cr}$) required for complete conversion of neutral particles to negatively charged particles for a fixed size of particle, an analytical asymptotic analysis is performed at the plateu regime of the population curve for negatively charged particles. As an example, the population dynamics curve for 500 nm size negatively charged particles at large time scale is fitted with asymptote: $N_{\rm neg} = 2.22\times 10^{-5}\times n_{p} + 0.7883$, where $N_{\rm neg}$ is the population of negatively charged particles and $n_p$ is the pulse number. To estimate the ``critical'' pulse number $n_{\rm cp}$ when the complete population conversion of neutral particles to negatively charged particle will occur is $N_{\rm neg} = 1 - N_{\rm pos}$ where the maximum value of positively charged particle population 0.01 (1$\%$) is considered. Now implementing these values in the above asymptotic equation, the approximate critical pulse number for a particular size of particle can be obtained. The large time scale asymptotic expressions are different for different size particles. It is to be noted that uncertainty is much higher to determine $n_{\rm cp}$ for particles $<$ 100 nm particles as the steady growth continues for the population curve of negatively charged particle within simulation time reported in this work. The dependence of critical pulse number on particle size is shown in Figure~\ref{critical_pulse_number_variation}. The bigger is the particle size, less number of pulses are needed for complete population conversion of neutral particles to negatively charged particles. On the other hand, for smaller size particles ($<$ 40 nm), the critical pulse number, $n_{\rm cr} \rightarrow \infty $.

In conclusion, the population dynamics of the charged particles inside and outside EUV beams have been considered for single and multi-pulse scenario for the first time. The populations of different species strongly depends on surrounding plasma conditions and associated plasma-particle interactions (outside EUV beam area) or plasma-particle-photon interactions (inside EUV path). At the beginning of single pulse during EUV-ON ($<$ 100 ns), the population of positively charged particles rises sharply due to particle-photon interactions, but subsequently fast decay occurs due to afterglow effect. However over multi-pulse scenario, the population of positively charged particles fluctuate but maintains a steady state average value. However, the population of negatively charged particles increases over pulses and that of neutral particles goes down. The critical pulse numbers have been identified for different size particles above which all the particles become negatively charged outside the beam locations. The population dynamics of positively charged particles strongly depends on beam energy. In the current investigation the beam energy is considered as 0.2 mJ. However with increasing beam energy (ex. beam energy = 1 mJ) the population dynamics of positively charged particles may change. Also it is very important to extend this study for nonspherical particles~\cite{Molotkov2000,Annaratone2001,Ivlev2003,Annaratone2003,Miloch2009,Okuzumi2009,Annaratone2011,ChaudhuriAPL2012,Matthews2012,Matthews2013,Matthews2014,Chaudhuri2016_q2d,Chaudhuri2016_di,Ivlev2020,Mengel2025} which can have high impact to assess realistic scenario on particle contamination. The detailed investigation of such studies are kept as future work. 

 Any data that support the findings of this study are included within the article.

* Corresponding author: manis.chaudhuri@asml.com

\end{document}